\documentclass[11pt, a4paper,oneside]{article}

\usepackage{ulem} 
\usepackage{graphicx}
\usepackage{multirow}

\begin{document}

% =========== BEGINNING OF THE PAPER ========================
\begin{center}
\textbf{VX Hya time keeping. Stroboscopic analysis}
\end{center}
\begin{center}
{\large{Bonnardeau, Michel}}
\\
MBCAA Observatory, Le Pavillon, 38930 Lalley, France \footnote{arzelier1@free.fr}
\end{center}

\abstract{Photometric observations of the double-mode pulsator VX Hya are presented. They are analyzed with a stroboscopic method, completed by Fourier analysis.}

\section{Introduction}\label{introduction}
VX Hydrae (RA=09h45min46.8s, DEC=-12°00'14", 2000.0) is a bright, high amplitude $\delta$ Scuti star, with its V magnitude between 10.1 and 10.9. As reported by Fitch, 1966 and Templeton et al, 2009, it is a double-mode pulsator with pulsations at the frequency $f_0=4.4765$ cycles/day (the fundamental) and  $f_1=5.7898$ cycles/day (the overtone), with almost the same amplitudes. There are many harmonics at frequencies $2f_0$, $3f_0$,... and $2f_1$, $3f_3$,... and also many beats at frequencies $f_0+f_1$, $f_1-f_0$,....
\\ \\
The pulsations themselves are variable: Bonnardeau et akl, 2011 reported a sudden change in 2008, and Xue et al, 2018 reported a decrease of the fundamental and overtone frequencies.
\\ \\
In this paper, new observations are reported. These data are completed with the observations of Fitch, 1966 and with AAVSO observations, to be analyzed with a stroboscopic method, completed by Fourier analysis. 
 
\section{Observations}\label{observations}
\begin{figure}[htbp]
	\centering
	\includegraphics [width=12cm]
	{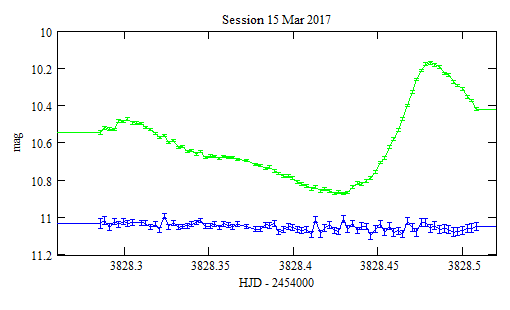}
	\caption{Green: the magnitudes of VX Hya, Blue: the magnitudes of a check star, GSC 5482-01054, shifted by -1.8 mag to fit into the graph. The error bars are the quadratic sums of the statistical uncertainty of the star and of the comparison.}
	\label{figA}
\end{figure}

My observations were carried out with a 203 mm f/6.3 Schmidt-Cassegrain telescope, a Johnson V filter, and a camera with a KAF401E CCD. I made time series with individual exposures of 200 s (a few measurements have 60 s).
\\ \\
For the differential photometry, the comparison star is TYC 5482-01347/1 with $V=11.580$, computed from the Tycho magnitudes owing to transformation formulas of Mamajek et al, 2002. According to the UCAC4/APASS catalog, it has $B-V=0.547$, while VX Hya has $B-V\approx0.40$.
\\ \\
I obtained 7922 measurements, in 126 sessions, from 2005 to 2021. The journal of observations is in Table~\ref{tabB}. An example of a light curve is in Fig.~\ref{figA}, and another one is in Fig.~\ref{figJ1}.

   %[htbp]}

\section{Fundamental and overtone frequencies determination}\label{frquencies}
I search for pulsations in my data plus those of Fitch, 1966. These last data consist of 1398 V-magnitude measurements, in 30 sessions, from 1954 to 1965. The comparison star is not the same, so I adjust  Fitch's magnitudes by adding -0.075  so that the average Fitch magnitude is equal to my average magnitude. Furthermore, the times are converted in BJD; between 1954 and 2021, this introduces a delay of up to 44 s, mostly due to the slowing down of Earth rotation (Eastman et al, 2010). The pulsations are then searched using the Period04 software program Lenz \& Breger, 2005. The fundamental and overtone pulsations are readily detected, but not the harmonics and beats on this 67-year long data set, indicating that they are not stable. The frequencies are:
\\ \\
$f^{all}_0=4.476,476,6(2)$ cycles/day for the fundamental \\
$f^{all}_1=5.789,769,1(3)$ cycles/day for the overtone \hspace*{\fill} (1)
\\ \\ 
The amplitude is 0.14 mag for the fundamental, and 0.11 mag for the overtone.
\\ \\
These frequency measurements are compatible with those of Xue et al, 2018, with a data set from 1955 (Fitch, 1966) to 2015.
\\ \\ 
Using only the Fitch, 1966 observations, the same analysis gives:
\\ \\
$f^{1954-65}_0=4.476,487(3)$  cycles/day\\
$f^{1954-65}_1=5.789,745(5)$  cycles/day \hspace*{\fill} 
\\ \\
and using only my observations:\\ \\
$f^{2004-21}_0=4.476,461,3(9)$ cycles/day\\
$f^{2004-21}_1=5.789,768(1)$ cycles/day  \hspace*{\fill}
\\ \\
The precision on these frequencies is good enough not to have any cycle lost, despite the large gap between the Fitch, 1966 data, before 1965, and the more recent ones, after 2003 (unless there were very large irregularities in the interval with no observation).	

\section{Stroboscopic analysis principle}\label{principle}
\begin{figure}[htbp]
	\centering
	\includegraphics [width=13.cm]  %13cm
	{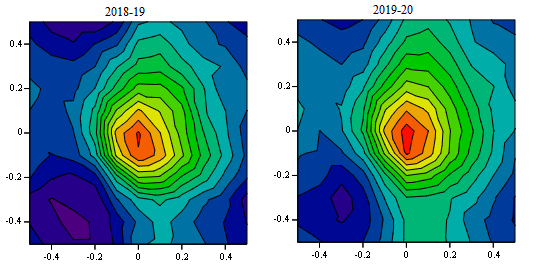}
	\caption{Stroboscopic phase plots using the frequencies of Table~\ref{tabD3}. The maxima are at the phases:
		$\varphi^{2018-19}_0=0.01\pm0.04$,
		$\varphi^{2018-19}_1=-0.02\pm0.055$,
		$\varphi^{2019-20}_0=0.00\pm0.05$,
		$\varphi^{2019-20}_1=-0.04\pm0.05$.}
	\label{figB}
\end{figure}

I analyze the pulsations, using a stroboscopic method:
\\ \\
VX Hya has two pulsations, the fundamental and the overtone, so the idea of a stroboscopic analysis by making 3D phase plots with the phase of the fundamental along the horizontal axis, the phase of the overtone along the vertical axis, and the magnitude as, for example, contour levels. When there are several magnitude measurements for a fundamental phase value and an overtone phase value, the average magnitude is used.
\\ \\
I make such stroboscopic phase plots, season per season, by calculating the phases as:
\\ \\
$\varphi_i=$fract$(t.f_i)$ 
\\ \\
with fract the fractional part, $i=0$ for the fundamental, $i=1$ for the overtone, $f_i$ the frequencies,  and $t$ the truncated BJD: TBJD=BJD-2454000. 
\\ \\
Such a stroboscopic analysis is sometimes done with cataclysmic variables as a function of the orbital and spin frequencies (see Littlefield et al, 2020). An example of stroboscopic phase plots for VX Hya is shown in Fig.~\ref{figB}, using the observations for the 2018-19 and the 2019-20 seasons, and the frequencies $f^{2016-21}_0$ and $f^{2016-21}_1$ derived from the Period04 analysis of the 2016-21 seasons, shown in Table~\ref{tabD3}.
\\ \\
With the use of contour levels from the stroboscopic phase plot, the maximum may be (slightly) shifted, due to the interpolation algorithm of the software package used to draw the contours\footnote{I use Mathcad from Parametric Technology Corporation.}. The positions of the maxima are then measured using the following iterative way:
\\ \\
step \textcircled{a}: the position of the maxima $\varphi^a_0\pm\Delta\varphi^a_0$,  $\varphi^a_1\pm\Delta\varphi^a_1$ is estimated by eye from the plot;
\\ \\
step \textcircled{b}: the data in the horizontal strip given by  $\varphi^a_1\pm\Delta\varphi^a_1$ are considered. The magnitudes as a function of $\varphi_0$ shows a bell shape whose maximum is measured, using a polynomial interpolation, and taking into account the uncertainties on the magnitudes and the dispersion of the data points, to give a new position $\varphi^b_0\pm\Delta\varphi^b_0$;
\\ \\
step \textcircled{c}: the data in the vertical strip given by  $\varphi^b_0\pm\Delta\varphi^b_0$ are considered. A new coordinate $\varphi^c_1\pm\Delta\varphi^c_1$ is derived the same way as in the previous step;
\\ \\
step \textcircled{d}: if $\varphi^c_1$ is significantly different from $\varphi^a_1$, step  \textcircled{b} is repeated using the strip given by $\varphi^c_1\pm\Delta\varphi^c_1$, and so on; actually, this converges quickly.
\\ \\
\noindent This stroboscopic method may be compared to the usual Fourier analysis. With the Fourier analysis, the magnitudes are decomposed as:
\\ \\
mag$(t)=Z+\sum\limits_{i} A_i\sin[2\pi(f_it+\Phi_i)]$
\\ \\
In the stroboscopic analysis, if the frequency is the same as in the Fourier decomposition, using the same TBJD, and with $A_i>0$, the maximum of brightness is at the stroboscopic phase:
\\ \\
$\varphi_i=\displaystyle\frac{3}{4}-\Phi_i$ 
   %[htbp]}
\\
\noindent
As an example, I analyze the 2016-21 time interval with the Period04 software program, using 5 seasons for a reliable determination of the frequencies. The results are shown in Table~\ref{tabD3}. The stroboscopic phase plots for the 2018-19 and 2019-20 seasons, using the frequencies $f^{2016-21}_0$ and $f^{2016-21}_1$ of Table~\ref{tabD3}, are shown in Fig.~\ref{figB}. The positions of the maxima are to be compared with the $\displaystyle\frac{3}{4}-\Phi_i$:
\\ \\
$\displaystyle\frac{3}{4}-\Phi^{2016-21}_0=0.081$ and $\displaystyle\frac{3}{4}-\Phi^{2016-21}_1=0.029$.
\\ \\
This is roughly in agreement, considering this is only an approximation because the harmonics and beats are not taken into account.

\section{Stroboscopic analysis results}\label{results}
Besides my observations from 2005 to 2021 and those of Fitch, 1966 from 1954 to 1965, I also use AAVSO observations for the 2002-4 (observer BIW) and 2005-6 (observer DKS) seasons. All the observations are done with a Johnson V filter. The comparison stars are not the same for the different observers (that does not matter as long as the data are not mixed up in one stroboscopic phase plot). The observations are available as the machine-readable file \textit{photometry.dat} in Appendix, with the observations of Fitch, 1966 obtained by OCR and my observations (the AAVSO observations may be downloaded from the AAVSO web site). The journal of observations is shown in Table~\ref{tabB}.

\begin{figure}[htbp]
\centering
\includegraphics [width=15cm]
{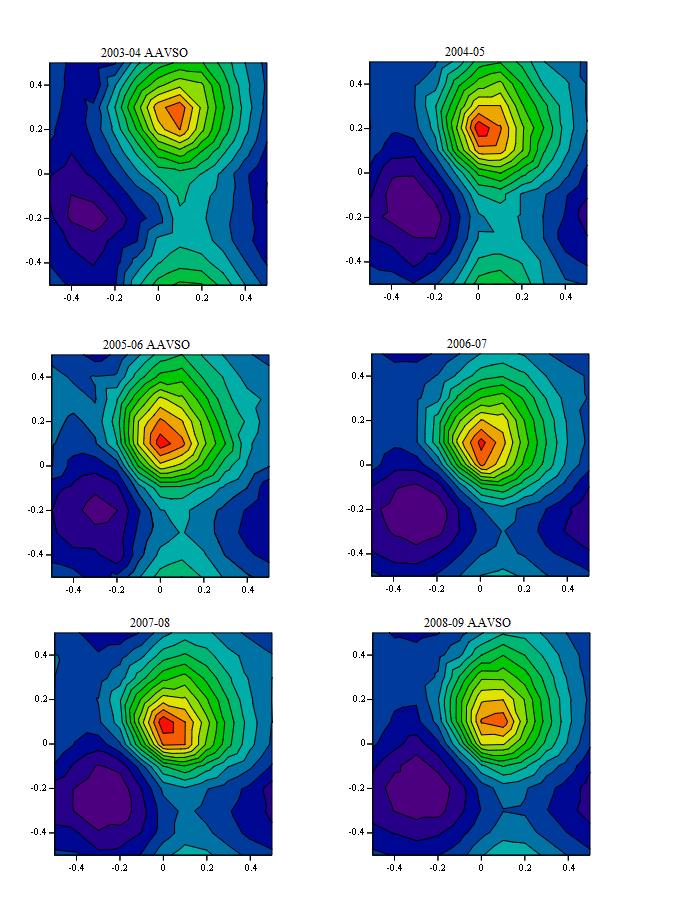}
\caption{2003-09 stroboscopic phase plots. }
\label{figF}
\end{figure}

\begin{figure}[htbp]
\centering
\includegraphics [width=15cm]
{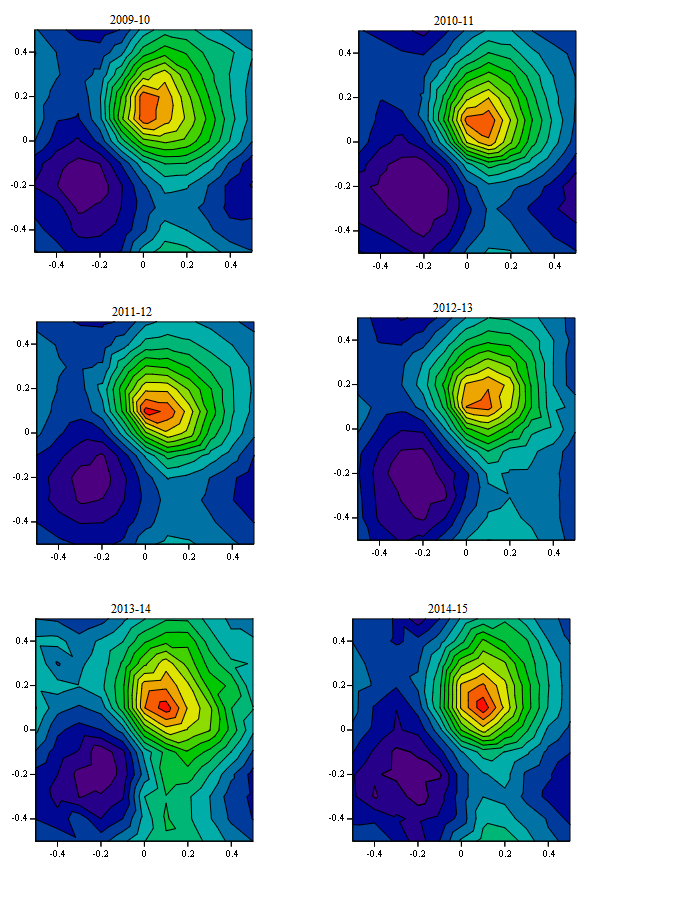}
\caption{2009-15 stroboscopic phase plots.}
\label{figG}
\end{figure}

\begin{figure}[htbp]
\centering
\includegraphics [width=15cm]
{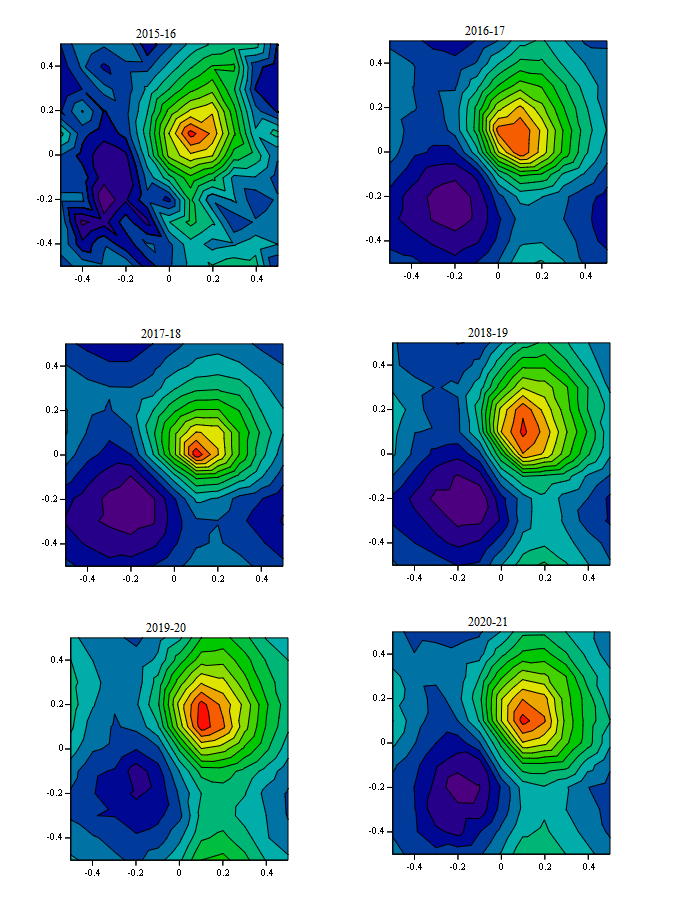}
\caption{2015-21 stroboscopic phase plots.}
\label{figH}
\end{figure}

\begin{table}[htbp]
	\centering
	\begin{tabular}{p{20mm}   p{20mm}  p{20mm}  p{22mm}}
			\hline
	Season &	Nb of & Nb of & Observer \\
	&  sessions & measurements 	& \\
	\hline	
	 1954-56 &		\multirow{2}{*} {10} & \multirow{2}{*}	{312} & \multirow{6}{*}{Fitch, 1966} \\
	 (2 seasons) & & & \\
	 1956-57 &	 	5 &	324 & \\
	 1957-58 &	 	5 &	302 & \\
	 1958-59 & 	5 &	350 & \\
	 1963-65& \multirow{2}{*}	5 & \multirow{2}{*} {110} & \\
		 (2 seasons) & & & \\
		 \hline
		 2002-03 &	4 &	263 &	\multirow{2}{*}{AAVSO BIW} \\
		 2003-04 & 	4 &	376 &  \\
		 \hline
		 2004-05 &	6 &	248 &	this work \\
		 \hline
		 2005-06 &	21 &	3098 &	AAVSO DKS \\
		 \hline
		 2006-07 &	4 &	162 &	\multirow{15}{*}	{this work} \\
		 2007-08 & 	7 &	346 & \\
		 2008-09 & 	9 &	267& \\
		 2009-10 & 	6 &	385& \\
		 2010-11 &	5& 	313& \\
		 2011-12 &	4& 	243& \\
		 2012-13& 	6 &	323& \\
		 2013-14 & 	11& 	644& \\
		 2014-15 & 	12 &	919& \\
		 2015-16 & 	5 &	404& \\
		 2016-17 & 	13 &	826& \\
		 2017-18 & 	4& 	251& \\
		 2018-19 & 	15 	&1211& \\
		 2019-20 &	12 &	845& \\
		 2020-21 & 7 & 535 & \\
		 \hline
	\end{tabular}
	\caption{Journal of observations.}
	\label{tabB}
\end{table}
\noindent The stroboscopic phase plots are made using the frequencies $f^{all}_i$ in (1). They are
shown in Fig.~\ref{figE}-Fig.~\ref{figH}, with roughly 50 mmag between each contour:
\begin{figure}[htbp]
	\centering
	\includegraphics [width=15cm]
	{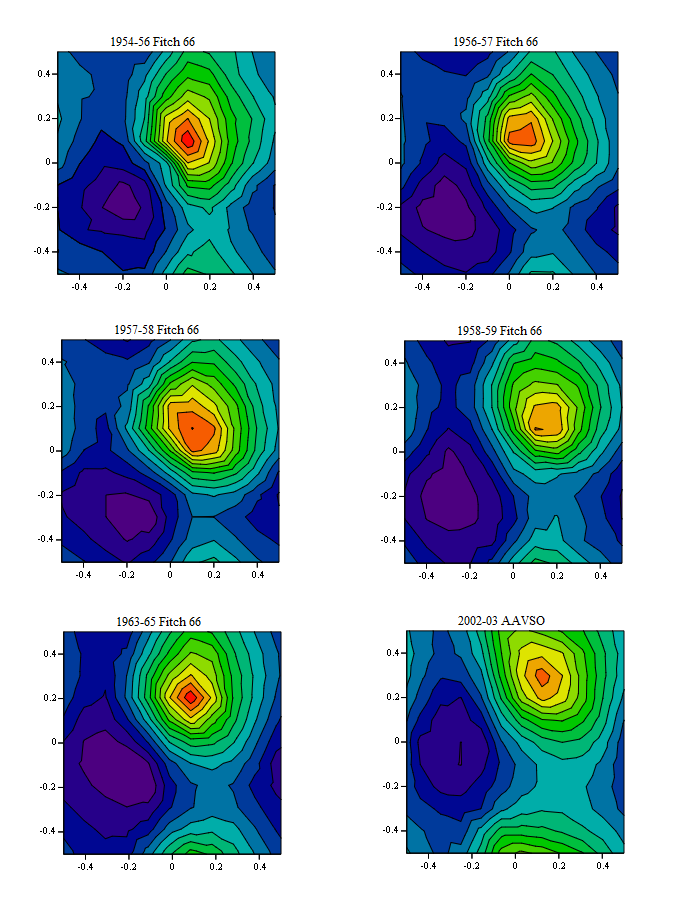}
	\caption{1954-2003 stroboscopic phase plots.}
	\label{figE}
\end{figure}  %[htbp]}

  %[htbp]}

  %[htbp]}

  %[htbp]}

\newpage

\noindent The phases of the brightness maxima are estimated by iterations, as explained in Section~\ref{principle}, and are listed in Table~\ref{tabC}, with $\varphi_0$ the phase for the fundamental (horizontal axis), $\varphi_1$ for the overtone (vertical axis).

\begin{table}[htbp]
	\centering
	\begin{tabular}{p{20mm}   p{20mm}  p{20mm}}
		\hline
Season & $\varphi_0$ & $\varphi_1$ \\
\hline
1954-56 &	$0.08\pm 0.04$ &	$0.15\pm0.06$ \\
1956-57 &	$0.025\pm0.03$ &	$0.14\pm0.05$ \\
1957-58 &$	0.09\pm0.05$ &	$0.06\pm0.06$ \\
1958-59 &	$0.14\pm0.03$ &$	0.10\pm0.05$ \\
1963-65 &	$0.08\pm0.03$& $	0.20\pm0.03$ \\
2002-03  &	$	0.13\pm0.03$& $ 	0.26\pm0.04$ \\
2003-04  &	$	0.01\pm0.06 $& $	0.25\pm0.03$ \\
2004-05  &	$	0.01\pm0.04 $& $	0.175\pm0.03$ \\
2005-06  &	$	0.025\pm0.055 $& $	0.105\pm0.045$ \\
2006-07  &	$	0.01\pm0.03 $& $	0.09\pm0.06$ \\
2007-08  &	$	0.04\pm0.04 $& $	0.07\pm0.04$ \\
2008-09  &	$	0.03\pm0.05 $& $	0.115\pm0.045$ \\
2009-10  &	$	0.00\pm0.02 $& $	0.15\pm0.06$ \\
2010-11  &	$	0.06\pm0.045 $& $	0.075\pm0.045$ \\
2011-12  &	$	0.07\pm0.04 $& $	0.105\pm0.035$ \\
2012-13  &	$	0.075\pm0.045 $& $	0.115\pm0.045$ \\
2013-14  &	$0.087\pm0.04 $& $	0.10\pm0.06$ \\
2014-15  &	$0.087\pm0.04 $& $	0.11\pm0.045$ \\
2015-16  &	$	0.11\pm0.05 $& $	0.065\pm0.035$ \\
2016-17  &	$	0.085\pm0.045 $& $ 	0.055\pm0.045$ \\
2017-18   &	$	0.055\pm0.06  $& $	0.085\pm0.06$ \\
2018-19   &	$	0.12\pm0.045  $& $	0.115\pm0.045$ \\
2019-20   &	$	0.125\pm0.04  $& $	0.14\pm0.055$ \\
2020-21 & $0.12\pm0.05$ & $0.105\pm0.045$ \\
\hline
\end{tabular}
\caption{Coordinates of the brightness maxima in the stroboscopic phase plots of Fig.~\ref{figE}-Fig.~\ref{figH}.}
\label{tabC}
\end{table}

\begin{figure}[htbp]
	\centering
	\includegraphics  [width=16cm]
	{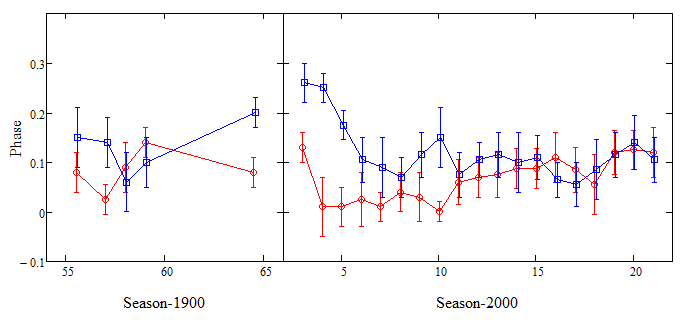}
	\caption{Maxima of the stroboscopic phase plots. Red: the phase for the fundamental pulsation, Blue: for the overtone. For the season axis, 55 means the season 1954-55 and so on.}
	\label{figJ}
\end{figure}

These phases of the brightness maxima of Table~\ref{tabC}  are plotted in Fig.~\ref{figJ}. 

\newpage
\section{Discussion}\label{discussion}
From the plot of Fig.~\ref{figJ} the phases appear to be fairly stable, except between 2003 and 2010; this was already noticed  and investigated by Bonnardeau et al, 2011, using a different method.
\\ \\
To further investigate how the pulsations may vary, I use the Period04 software program over several consecutive seasons. The number of seasons must be large enough to have reliable measurements of the frequencies, but not too large for the pulsations not to be scrambled by variations of frequencies. I use the Monte Carlo feature of Period04 to evaluate the quality of the fits. This analysis is done on intervals of 5 seasons: 1954-59, 2010-15, 2016-21 (there is no good fit in the 2003-10 interval). The results, up to the third harmonics, are shown in Table~\ref{tabD1},  Table~\ref{tabD2}, and  Table~\ref{tabD3}.

\begin{table}[htbp]
	\centering
	\begin{tabular}{p{48mm} p{28mm} p{28mm}}
		\hline
	Frequency 	$f^{1955-59}$ & Amplitude & Phase $\Phi^{1955-59}$\\
		(cycles/day) & $A^{1955-59}$ (mag) &	\\
		\hline
			$f^{1955-59}_0=4.4764966(46)$ &	 0.1483(18) &	 0.9265(17) \\
				$f^{1955-59}_15.7897618(52)$ &	 0.1204(25) 	& 0.4168(25) \\
			$f^{1955-59}_0+f^{1955-59}_1$ &	 0.0620(19) & 	 0.5105(59) \\
			$2f^{1955-59}_0$ &	 0.0433(19) &	 0.6327(68) \\
			$f^{1955-59}_1-f^{1955-59}_0$ &	 0.0335(19) &	 0.4426(72) \\
			$2f^{1955-59}_1$	& 0.0293(18) &	 0.601(53) \\
				$f^{1955-59}_0+2f^{1955-59}_1$ &	 0.0183(15) &	 0.455(17) \\
		$2f^{1955-59}_0-f^{1955-59}_1$  &	 0.0133(22) &	 0.818(35) \\
		$2f^{1955-59}_0+f^{1955-59}_1$	& 0.0126(17) &	 0.775(24) \\
		$3f^{1955-59}_0$	& 0.0075(16) &	 0.635(34) \\
			$2f^{1955-59}_1-f^{1955-59}_0$ &	 0.0104(39) &	 0.73(37) \\
		$3f^{1955-59}_1$	& 0.0064(17) &	 0.715(43) \\
		\hline 
	\end{tabular}
\caption{The Fourier decomposition for the 1955-59 time interval (5 seasons), with (average and standard deviation) $<t>^{1955-59}=-17992\pm485$ TBJD. The zeropoint is $Z=10.638$ mag and the residuals $\chi=29$ mmag.}
\label{tabD1}
\end{table}

\begin{figure}[htbp]
	\centering
	\includegraphics  [width=12cm]
	{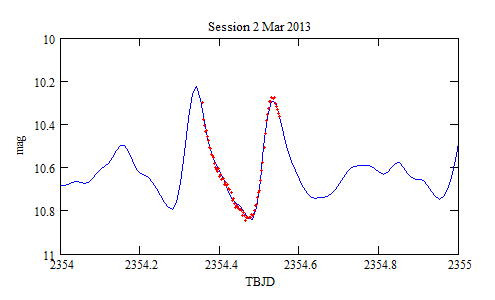}
	\caption{Blue line: the model given in Table~\ref{tabD2}, Red dots: the observations.}
	\label{figJ1}
\end{figure}

\begin{table}[htbp]
	\centering
	\begin{tabular}{p{48mm} p{28mm} p{28mm}}
		\hline
	Frequency 	$f^{2010-15}$ & Amplitude & Phase $\Phi^{2010-15}$\\
		(cycles/day) & $A^{2010-15}$ (mag) &	\\
		\hline
			 $f^{2010-15}_0=4.4764589(29)$ &	 0.1392(17) &	 0.6391(22) \\
			$f^{2010-15}_1=5.7897436(44)$ &	 0.1134(15) &	 0.6182(27) \\
		$f^{2010-15}_0+f^{2010-15}_1$ &	 0.0525(21) &	 0.6234(46) \\
		$2f^{2010-15}_0$ &	 0.041(12) &	 0.67(28) \\
		$f^{2010-15}_1-f^{2010-15}_0$	& 0.0298(83) &	 0.46(24) \\
		$2f^{2010-15}_1$	& 0.0235(19) &	 0.723(13) \\
			$f^{2010-15}_0+2f^{2010-15}_1$	& 0.0164(14) &	 0.638(17) \\
		 $2f^{2010-15}_0-f^{2010-15}_1$ &	 0.0108(17) &	 0.869(23) \\
		$2f^{2010-15}_0+f^{2010-15}_1$ &	 0.0135(47) &	 0.70(32) \\
		$3f^{2010-15}_0$ &	 0.0095(14) &	 0.808(25) \\
	$2f^{2010-15}_1-f^{2010-15}_0$	& 0.0090(33) &	 0.55(26) \\
		$3f^{2010-15}_1$	& 0.0066(24) &	 0.60(45) \\
		\hline 
	\end{tabular}
\caption{The Fourier decomposition for the 2010-15 time interval (5 seasons), with (average and standard deviation) $<t>^{2010-15}=2583\pm505$ TBJD. The zeropoint is $Z=10.613$ mag and the residuals $\chi=36$ mmag.}
\label{tabD2}
\end{table}
\newpage
\begin{table}[htbp]
	\centering
	\begin{tabular}{p{48mm} p{28mm} p{28mm}}
		\hline
		Frequency  $f^{2016-21}$ & Amplitude & Phase $\Phi^{2016-21}$\\
		(cycles/day) & $A^{2016-21}$ (mag) &	\\
		\hline
 $f^{2016-21}_0=4.4764534(18)$ &	 0.14295(81) &	 0.6689(10) \\
 $f^{2016-21}_1=5.7897347(28)$ &	 0.11363(97) &	 0.7212(12) \\
	$f^{2016-21}_0+f^{2016-21}_1$ &	 0.05418(84) &	 0.7738(28) \\
 $2f^{2016-21}_0$ &	 0.03974(87) &	 0.7056(38) \\
 $f^{2016-21}_1-f^{2016-21}_0$ &	 0.02810(93) &	 0.9149(54) \\
	$2f^{2016-21}_1$ &	 0.0250(10) &	 0.9093(55) \\
	$f^{2016-21}_0+2f^{2016-21}_1$	& 0.0160(57) &	 0.87(12) \\
 $2f^{2016-21}_0-f^{2016-21}_1$ &	 0.01265(90) &	 0.559(11) \\
	$2f^{2016-21}_0+f^{2016-21}_1$ &	 0.01322(97) &	 0.868(13) \\
	$3f^{2016-21}_0$	& 0.0109(43) &	 0.62(21) \\
 $2f^{2016-21}_1-f^{2016-21}_0$ &	 0.00795(86) &	 0.037(17) \\
 $3f^{2016-21}_1$ &	 0.0066(26) &	 0.94(28) \\
\hline 
\end{tabular}
\caption{The Fourier decomposition for the 2016-21 time interval (5 seasons), with (average and standard deviation) $<t>^{2016-21}=4542\pm505$ TBJD. The zeropoint is $Z=10.637$ mag and the residuals $\chi=30$ mmag.}
\label{tabD3}
\end{table}
\noindent An example of an observed light curve fitted with the Fourier model is shown Fig.~\ref{figJ1}.
\\ \\
\noindent From Table~\ref{tabD1}, Table~\ref{tabD2}, and Table~\ref{tabD3}, it appears that the frequencies $f_0$ and $f_1$ are decreasing. The best fits yield the derivatives:
\\ \\
$\dot{f}_0=-19.1\pm1.1*10^{-10}$ cycles/d$^2$
\\ 
$\dot{f}_1=-10.7\pm1.9*10^{-10}$ cycles/d$^2$
\\ \\
The value found for the derivative of the fundamental frequency $\dot{f}_0$ is about the same as the one found by Xue et al, 2018, using a different method. But for the overtone, the frequency derivative value found, comparable to the one of the fundamental, is stronger than the upper limit of Xue et al, 2018.

\setcounter{secnumdepth}{0}
\section*{Acknowledgements}
	The use of the AAVSO International Database is acknowledged, especially the observations of Butterworth N. (code BIW) and Dvorak S. (code DKS).
\\ \\
The use of the online tool of the University of Ohio to convert HJD to BJD, at \\
http://astroutils.astronomy.ohio-state.edu/time/hjd2bjd.html, is acknowledged. 
	
\section*{References}

% tilde in URLs must be given this way to work correctly with hyperref
Bonnardeau M., Dvorak S., Poklar R., Samolyk G., 2011, JAAVSO \textbf{39} 1.
\\ \\
Eastman J., Siverd R., Scott Gaudi B., 2010, PASP \textbf{122} 935.
\\ \\
Fitch W.S., 1966, ApJ \textbf{143} 852.
\\ \\
Lenz P., Breger M., 2005, Comm. in Asteroseismology \textbf{146} 53.
\\ \\
Littlefield C., Garnavich P., Kennedy M.R. et al., 2020, ApJ \textbf{896} 116.
\\ \\
Mamajek E.E., Meyer M.R., Liebert J., 2002, Astronom. J. \textbf{124} 1650
and 2006, Astronom. J. \textbf{131} 2360.
\\ \\
Templeton M.R., Samolyk G., Dvorak S., Poklar R., Butterworth N., Gerner H., 2009, PASP \textbf{121} 1076.
\\ \\
Xue H.F., Fu J.N., Fox-Machado L. et al., 2018, ApJ \textbf{861} 96.

\end{document}